\documentclass[reprint,prl,twocolumn,superscriptaddress,showpacs,showkeys,floatfix,preprintnumbers]{revtex4-2}
\usepackage{braket}
\usepackage{booktabs}
\usepackage{xargs}
\usepackage{mathtools}
\usepackage[english]{babel}
\usepackage[utf8]{inputenc}
\usepackage{amsmath,amssymb,braket,slashed,mathrsfs}
\usepackage{microtype}
\usepackage{pifont}
\usepackage{MnSymbol}
\usepackage{dcolumn}
\usepackage{graphicx}
\usepackage{tikz}
\usetikzlibrary{shapes,arrows,positioning}
\tikzset{decision/.style={diamond, draw, fill=blue!20, text width=4.5em, text badly centered, inner sep=0pt}}
\tikzset{block/.style={rectangle, draw, fill=blue!20, text width=10em, text centered, rounded corners, minimum width=3.5cm}}
\tikzset{block1/.style={rectangle, draw, fill=blue!20, text width=18.5em, text centered, rounded corners, minimum width=3.5cm}}
\tikzset{line/.style={draw, -latex, thick}}
\usepackage[colorlinks=true, allcolors=blue]{hyperref}
\usepackage{multirow,array}
\usepackage{physics}
\usepackage{orcidlink}

\usepackage{cleveref}
\newcommand{\be}{\begin{equation}}
	\newcommand{\ee}{\end{equation}}
\newcommand{\ba}{\begin{eqnarray}}
\newcommand{\ea}{\end{eqnarray}}

\renewcommand{\vec}[1]{\mathbf{#1}}

\newcommand{\affA}{Department of Physics and Astronomy, University of Tennessee, Knoxville, TN 37996, USA}
\newcommand{\affB}{Physics Division,
  Oak Ridge National Laboratory, Oak Ridge, TN 37831, USA}
\newcommand{\affC}{National Center for Computational Sciences, Oak Ridge National Laboratory, Oak Ridge, TN 37831, USA}
\newcommand{\affD}{Technion – Israel Institute of Technology, Haifa, Israel}
\newcommand{\affE}{PRISMA$^{++}$ Cluster of Excellence, Institut f\"ur Kernphysik, Johannes Gutenberg-Universit\"at, Mainz, Germany}

\begin{document}
\title{Taming nuclear size and shape effects in superallowed $\beta$-decay}
%\title{Combined analysis of superallowed $\beta $ decay rates with \textit{ab initio} calculations and experiment}

\author{Bingcheng He\orcidlink{0000-0003-4483-1992}}\affiliation{\affA}
\author{Mikhail Gorchtein\orcidlink{0000-0001-9348-0557}}\affiliation{\affE}
\author{Matthias Heinz\,\orcidlink{0000-0002-6363-0056}}\affiliation{\affC}\affiliation{\affB}
\author{Ben Ohayon\orcidlink{0000-0003-0045-5534}}\affiliation{\affD}
\author{Lucas Platter\orcidlink{0000-0001-6632-8250}}\affiliation{\affA}\affiliation{\affB}
\author{Chien-Yeah Seng\orcidlink{0000-0002-3062-0118}}\affiliation{\affA}

\date{\today}

\begin{abstract}

We present the first combined analysis of the statistical rate function $f$ in superallowed $\beta$ decays with \textit{ab initio} calculations and data. We focus on ${}^{10}\text{C}\rightarrow {}^{10}\text{B}^*$, ${}^{14}\text{O}\rightarrow {}^{14}\text{N}^*$ and ${}^{26\text{m}}\text{Al}\rightarrow {}^{26}\text{Mg}$, all of which are important channels for the precise determination of the Cabibbo-Kobayashi-Maskawa (CKM) matrix element $V_{ud}$. Nuclear charge form factors are obtained by combining experimental data on nuclear charge radii and theory calculations of ratios of moments with the in-medium similarity renormalization group, while the $\beta$ decay form factors are derived from exact isospin relations. This enables a rigorous study of the nuclear shape dependence in the statistical rate function $f$ and the quantification of its uncertainties from both experiment and theory. The calculation leads to a more precise test for the first-row CKM unitarity with reduced theoretical uncertainties. This work demonstrates a reliable strategy for combining nuclear many-body calculations with high-precision nuclear data to describe $\beta$ decays at tree level for precision tests of the Standard Model.

\end{abstract}
	
\maketitle

%\section{Introduction}

\textbf{Introduction} -- Superallowed $0^+ \rightarrow 0^+$ nuclear $\beta$ decays are central to precision tests of the Standard Model (SM)~\cite{GLASHOW1961579,Salam1959,PhysRevLett.19.1264}, providing the most precise determination of the Cabibbo-Kobayashi-Maskawa (CKM)~\cite{Cabibbo:1963yz,Kobayashi:1973fv} matrix element $V_{ud}$. CKM unitarity at the first row implies the vanishing of the quantity
\begin{equation}
\Delta_\text{CKM}\equiv|V_{ud}|^2 + |V_{us}|^2 + |V_{ub}|^2 -1~,
\end{equation}
which is currently tested at the sub-permille level~\cite{Hardy:2020qwl}. Any significant deviation of $\Delta_\text{CKM}$ from zero would signal physics beyond the SM. Great interest has been triggered, in particular, by the observation of a $\sim 3\sigma$ deviation of the first-row unitarity through a global fit of $V_{ud}$ and $V_{us}$ obtained from nuclear, 
neutron, 
pion, and kaon decays~\cite{Cirigliano:2022yyo}. Moreover, the agreement of $V_{ud}$ values from individual superallowed transitions constrains possible scalar interactions via the Fierz interference term~\cite{Jackson:1957zz,Hardy:2004id,Hardy:2004dm}.

A sub-permille extraction of $V_{ud}$ from superallowed $\beta$ decays requires precise knowledge of the $\mathcal{F}t$ value~\cite{Hardy:2020qwl},
\begin{equation}
    \mathcal{F}t = ft(1+\delta_R')(1+\delta_{\rm NS}-\delta_C)~,
\end{equation}
where $t$ is the partial half-life and $f$, $\delta_R'$, $\delta_{\rm NS}$, and $\delta_C$ encode nuclear-structure-dependent effects. Here we focus on the statistical rate function~\cite{Hardy:2004id,Hardy:2008gy}
\begin{equation}
    f=m_e^{-5}\int_{m_e}^{E_0}pE(E_0-E)^2F(E)C(E)Q(E)R(E)r(E)\,dE~,\label{eq:f}
\end{equation}
the phase-space integral of the $\beta$ spectrum that includes the Fermi function $F$, shape factor $C$, atomic screening $Q$, kinematic recoil $R$, and atomic overlap $r$. The factors $F$ and $C$ depend on the nuclear size through the charge density $\rho_\text{ch}(r)$ and the transition density $\rho_\text{cw}(r)$, respectively. Earlier studies considered them as unrelated quantities using empirical parameterizations and the nuclear shell model~\cite{Hardy:2004id} or treated $\rho_\text{cw}$ perturbatively~\cite{Wilkinson:1993hx,Hayen:2017pwg}. Recent studies uncovered isospin relations among them~\cite{Seng:2022epj,Seng:2022inj}:
\begin{eqnarray}
\rho_\text{cw}&=&\rho_\text{ch}^1+Z_0(\rho_\text{ch}^0-\rho_\text{ch}^1)=\rho_\text{ch}^1+(Z_{-1}/2)(\rho_\text{ch}^{-1}-\rho_\text{ch}^1)\nonumber\\
2Z_0\rho_\text{ch}^0&=&Z_1\rho_\text{ch}^1+Z_{-1}\rho_\text{ch}^{-1}~,\label{eq:isospin}
\end{eqnarray}
where the superscript of $\rho_\text{ch}$ and the subscript of the nuclear charge $Z$ denotes the third isospin component $T_z$ (with $T_z=1/2$ for neutrons). This enables a data-driven treatment of nuclear size effects in $f$ using measured charge radii. A reanalysis of the ${}^{26\mathrm{m}}\text{Al}\rightarrow {}^{26}\text{Mg}$ transition partially restored the first-row CKM unitarity~\cite{Gorchtein:2025wli}.

%The aforementioned procedure is limited by the available information on nuclear charge densities. 
High-precision charge radii from muonic atom and isotope shift measurements exist for only a small subset of superallowed emitters~\cite{Angeli:2013epw}. Moreover, beyond the radius, $f$ depends on the detailed shape of the charge density $\rho_{\text{ch}}(r)$ encoded in the nuclear form factor at finite momentum transfer $q^2$. This shape information usually comes from electron-nucleus scattering data, which are lacking for nearly all unstable nuclei (with recent progress at radioactive-ion-beam facilities~\cite{SCRIT}) and for many stable daughter nuclei~\cite{DeVries:1987atn}. It has been common to infer the density shape from the nearest stable isotope~\cite{Hardy:2004id}, introducing an additional uncontrolled approximation. As a result, the data-driven re-analysis of $f$ could only be applied to 15 of the 23 measured superallowed transitions and carries substantial uncertainty from the assumed density shape~\cite{Seng:2023cgl}.

Recent advances in \textit{ab initio} nuclear theory offer a path forward. Chiral effective field theory (EFT) provides a systematic, order-by-order improvable framework for low-energy nuclear interactions~\cite{Epelbaum2009,Machleidt2011}. Many-body calculations based on chiral EFT Hamiltonians have progressed rapidly due to methodological developments and growing computational power~\cite{Hergert2020}, enabling \textit{ab initio} studies from light to heavy nuclei, including both closed- and open-shell systems~\cite{King:2020wmp, Stroberg2021PRL_AbInitioLimits, Hu2022, Frosini:2021sxj,  Hebeler2023, Belley2024lec, Elhatisari2024N_WaveFunctionMatching, Tichai2024PLB_BCC, Sun2025PRX_Multiscale, Door2025PRL_YbBoson, bonaiti2025structuredoublymagicnuclei, Ng:2025hgx}. In particular, the in-medium similarity renormalization group (IMSRG)~\cite{Hergert2016,Stroberg2019} has become a widely used \textit{ab initio} method for nuclear-structure studies, including densities and form factors~\cite{EM7.5, HEINZ2025139975, MIYAGI2026140032, Hijikata:2026xwa}.

In this Letter, we establish a novel prescription for a high-precision, model-independent determination of $f$ by combining experimental information on nuclear charge densities with \textit{ab initio} calculations. The method applies to any superallowed decay for which at least one charge radius in the isotriplet is experimentally known. We illustrate its impact for ${}^{10}\text{C}\!\rightarrow{}^{10}\text{B}^*$, ${}^{14}\text{O}\!\rightarrow{}^{14}\text{N}^*$, and ${}^{26\mathrm{m}}\text{Al}\!\rightarrow{}^{26}\text{Mg}$: the first two are especially sensitive to new scalar interactions, while the third yields the most precise $V_{ud}$. For these cases, the \textit{ab initio} determination of moment ratios reduces uncertainties in $f$ from nuclear-size effects beyond the charge radii to below the $0.01\%$ level. We provide a rigorous uncertainty analysis, discuss consequences for $V_{ud}$ and CKM unitarity, and outline future opportunities, including connections to atomic-physics inputs and experimental efforts.

\textbf{Model independence of $f$} -- Determining $f$ requires the Fermi function $F(E)$, obtained by solving the Dirac equation for the outgoing positron in the static Coulomb potential generated by the charge density of the daughter nucleus, and the shape factor $C(E)$, computed from a charged weak density $\rho_{\text{cw}}$. Using isospin symmetry [Eq.~\eqref{eq:isospin}], $\rho_{\text{cw}}$ can be fixed if an additional charge density in the isotriplet is known. Although the charge density can be obtained from a Fourier transform of the charge form factor $F_{\text{ch}}(q^2)$, it is highly sensitive to the extrapolation of $F_{\text{ch}}$ to large $q^2$~\cite{Noel:2024led}, a region scarcely constrained by experiment and less reliably predicted by theory. Different large-$q^2$ extrapolations therefore imply different charge-density models, raising the key issue of the model dependence of $f$.

\begin{figure}
	\centering
	\includegraphics[width=0.95\columnwidth]
    {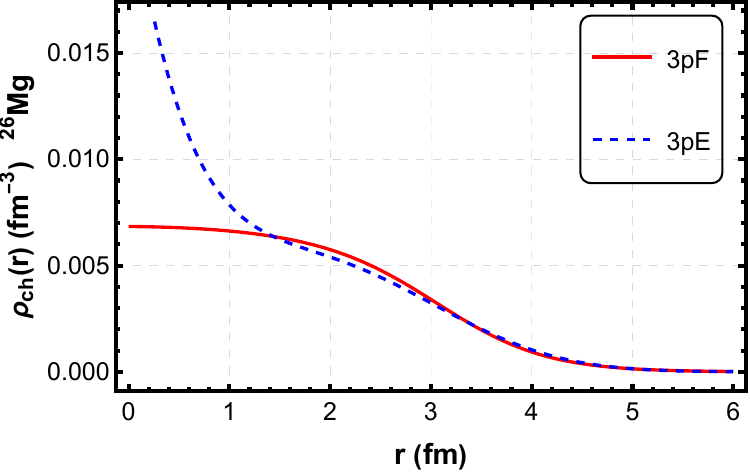}
	\caption{Nuclear charge density for ${}^{26}\text{Mg}$ in two different models with the first three moments fixed: $\langle r^2\rangle=9.1809\:\text{fm}^2$, $\langle r^4\rangle=125.521\:\text{fm}^4$, $\langle r^6\rangle=2204.11\:\text{fm}^6$. }
	\label{fig:densities}
\end{figure}

Recently, Ref.~\cite{Yerokhin:2025msa} showed that in highly-charged ions, fixing the three lowest moments $\langle r^{2,4,6}\rangle$ makes observables largely insensitive to the detailed charge-density model: even very different densities can reproduce identical energy levels up to $10^{-5}$ accuracy. This allows for a model-independent extraction of nuclear charge radii from Li-like ions. We find that the same holds for the statistical rate function $f$. To illustrate this, we compare the standard three-parameter Fermi (3pF) charge density with an intentionally unrealistic three-parameter exponential (3pE) model (see Ref.~\cite{Yerokhin:2025msa} for definitions). Although Fig.~\ref{fig:densities} shows that the two densities differ strongly near the nuclear center while sharing the same $\langle r^{2,4,6}\rangle$, the resulting $f$ values for the lightest (${}^{10}\text{C}$) and heaviest (${}^{26\mathrm{m}}\text{Al}$) cases studied are nearly identical:
\begin{eqnarray}
{}^{10}\text{C}\rightarrow {}^{10}\text{B}^*&:&2.30160\:(\text{3pF})\:\mathrm{vs.}\:2.30161\:(\text{3pE}),\nonumber\\
{}^{26\mathrm{m}}\text{Al}\rightarrow{}^{26}\text{Mg}&:&478.027\:(\text{3pF})\:\mathrm{vs.}\:478.034\:(\text{3pE}). \nonumber
\end{eqnarray}
The differences are $\le 0.001\%$, well below our precision goal of $0.01\%$ for $V_{ud}$, demonstrating that once $\langle r^{2,4,6}\rangle$ are fixed, $f$ is essentially model independent. Consequently, the required nuclear-size input is reduced from the full density to its three lowest moments, which we determine precisely below by combining experimental information with \textit{ab initio} calculations.

\textbf{\textit{Ab initio} calculations} -- We use the valence-space IMSRG (VS-IMSRG)~\cite{Hergert2016,Stroberg2019} to solve the many-body Schrödinger equation for nuclear Hamiltonians with two- and three-nucleon forces from chiral effective field theory. We consider two truncations of the VS-IMSRG equations: the VS-IMSRG(2), truncated at the normal-ordered two-body level, and the VS-IMSRG(3f$_2$), which approximately captures the leading effects of normal-ordered three-body operators~\cite{PhysRevC.110.044317}. We use three nuclear Hamiltonians: 1.8/2.0~(EM)~\cite{Hebeler2011}, 1.8/2.0~(EM7.5)~\cite{EM7.5}, and $\Delta$NNLO$_\mathrm{GO}$ (with cutoff 394 MeV)~\cite{N2LOgo}.
This allows us to explore and quantify the uncertainty associated with the input nuclear interactions.

To compute charge-density moments, one can either construct the corresponding moment operators and evaluate them directly~\cite{Hagen2016, Door2025PRL_YbBoson}, or, more conveniently, extract moments $\langle r^{2n}\rangle$ from derivatives of the charge form factor at low momentum transfer~\cite{MIYAGI2026140032}. The operator approach becomes cumbersome for higher moments~\cite{10.3389/fphy.2025.1581854,10.1093/ptep/ptz121, PhysRevC.104.024316} and treats translational invariance only approximately due to particle-rank truncations~\cite{Door2025PRL_YbBoson}. Here we therefore compute $F_{\mathrm{ch}}(q^2)$ with the VS-IMSRG at discrete low-$q^2$ points ($<1~\text{fm}^{-2}$) using
\begin{align}
F_{\mathrm{ch}}(q^2) &= e \sum_{i=1}^{A} \left\{ G_E^i(q^2)\left(1-\frac{q^2}{8m^2}\right) j_0(qr_i) \right. \nonumber \\
&\quad- \left. \frac{q^2}{2m^2}\left[G_M^i(q^2)-\frac{1}{2}G_E^i(q^2)\right]
(\vec{\ell}_i\cdot\vec{\sigma}_i)\frac{j_1(qr_i)}{qr_i} \right\},\label{eq:Fch}
\end{align}
where $q$ is the momentum transfer, $e$ the elementary charge, $m$ the nucleon mass, $G_{E,M}^i$ the nucleon electric/magnetic form factors~\cite{YE20188}, $\vec{\ell}_i$ and $\vec{\sigma}_i$ the orbital-angular-momentum and spin operators, and $j_{0,1}$ spherical Bessel functions. The expression above accounts for nucleon-size, Darwin-Foldy, and spin-orbit corrections~\cite{Krebs2020, MIYAGI2026140032}. To remove center-of-mass contamination and restore translational invariance, we correct the form factor using the Gaussian factorization of the center-of-mass wave function, established in coupled-cluster~\cite{PhysRevLett.103.062503} and IMSRG calculations~\cite{HEINZ2025139975}:
\begin{equation}
F_{\mathrm{ch}}^{\mathrm{int}}(q^2) = e^{q^2 b_{\mathrm{cm}}^2/4} F_{\mathrm{ch}}(q^2),\label{eq:Fchint}
\end{equation}
where \(F_{\mathrm{ch}}^{\mathrm{int}}(q^2)\) is the intrinsic charge form factor, \(F_{\mathrm{ch}}(q^2)\) is the calculated form factor that includes center-of-mass contamination, and \(b_{\mathrm{cm}}\) is the oscillator length associated with center-of-mass motion, computed from the expectation value of the center-of-mass kinetic energy.

%\subsection{Statistical rate function}

\begin{figure}
	\centering
	\includegraphics[width=0.9\columnwidth]
    {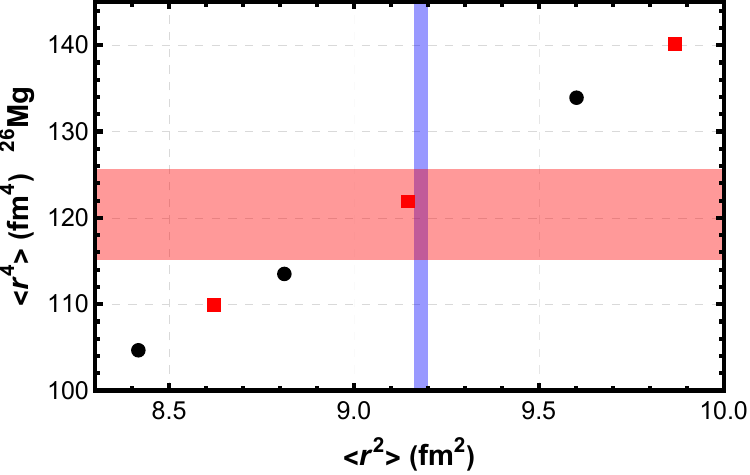}
    \includegraphics[width=0.9\columnwidth]
    {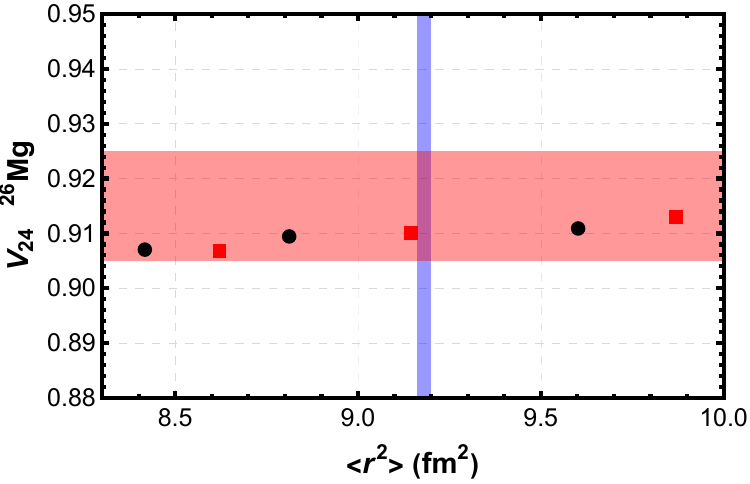}
	\caption{
    \textit{Ab initio} predictions for $\langle r^2\rangle$ vs.\ $\langle r^4\rangle$ (upper panel) and $\langle r^2\rangle$ vs.\ $V_{24}$ (lower panel) for ${}^{26}\text{Mg}$.
    Black dots indicate VS-IMSRG(2) results for three chiral EFT Hamiltonains, and red dots indicate VS-IMSRG(3f$_2$).
    Vertical and horizontal bands are experimental results obtained from muonic atom~\cite{Ohayon:2024dwt} and electron scattering~\cite{Soundranayagam:1988vdx}, respectively. Uncertainty bands for the scattering results are derived from how the calculations converge as momentum transfer increases.}
	\label{fig:Mgmoments}
\end{figure}

%The moments of $\rho_\text{ch}(q^2)$ can be obtained as:
%\begin{equation}
%    \langle r^{2n}\rangle=\frac{(-1)^n(2n+1)!}{n!}\frac{d^n \rho(q^2)}{d(q^2)^n}~.
%\end{equation}
%Derivatives can be obtained from discrete points of $\rho(q^2)$ through Gaussian processes. 

To assess theoretical accuracy, we compare the \textit{ab initio} moment predictions with experiment. The top panel of Fig.~\ref{fig:Mgmoments} shows that VS-IMSRG results for absolute moments $\langle r^{2,4}\rangle$ depend strongly on the chosen Hamiltonian and truncation of the many-body method, implying sizable absolute uncertainties that are not competitive with experimental precision, particularly for $\langle r^2\rangle$ from muonic atoms. However, these uncertainties are correlated between moments~\cite{Yerokhin:2025msa}, so the VS-IMSRG can predict the dimensionless moment ratios (``$V$-factors'')~\cite{Fricke:1995zz}
\begin{equation}
V_{2\:2n}\equiv\langle r^2\rangle^{1/2}/\langle r^{2n}\rangle^{1/2n}\,,
\end{equation}
with high precision. This is evident in the lower panel of Fig.~\ref{fig:Mgmoments}, where the calculated $V_{24}$ in ${}^{26}\text{Mg}$ are fully consistent with electron-scattering results, but with a much smaller uncertainty. Therefore, the optimal strategy is to combine $\langle r^2\rangle$ from muonic atoms and the $V$-factors from \textit{ab initio} calculations. This allows us to obtain the moments $\langle r^{2,4,6}\rangle$ with maximum precision and minimize the density-shape related uncertainty of $f$. 

This procedure also explains why we compute only the charge density $\rho_\text{ch}$ (rather than the charged weak density $\rho_\text{cw}$) with \textit{ab initio} methods. A reliable determination of $\rho_\text{cw}$ requires external input for the mean square ``charged weak radius'' $\langle r^2\rangle_{\text{cw}}$, which can only be extracted from high-precision measurements of nuclear-recoil effects in $\beta$ decays (as in the BeEST experiment~\cite{Leach:2021bvh}); such measurements are not yet available for superallowed decays. Although one could infer $\langle r^2\rangle_\text{cw}$ from charge radii using the isospin relation~\ref{eq:isospin}, it would not change our current strategy, which is to first deduce $\rho_\text{ch}$ with \textit{ab initio} methods and data and then obtain $\rho_\text{cw}$ using isospin symmetry.

\begin{figure}
	\centering
	\includegraphics[width=0.95\columnwidth]{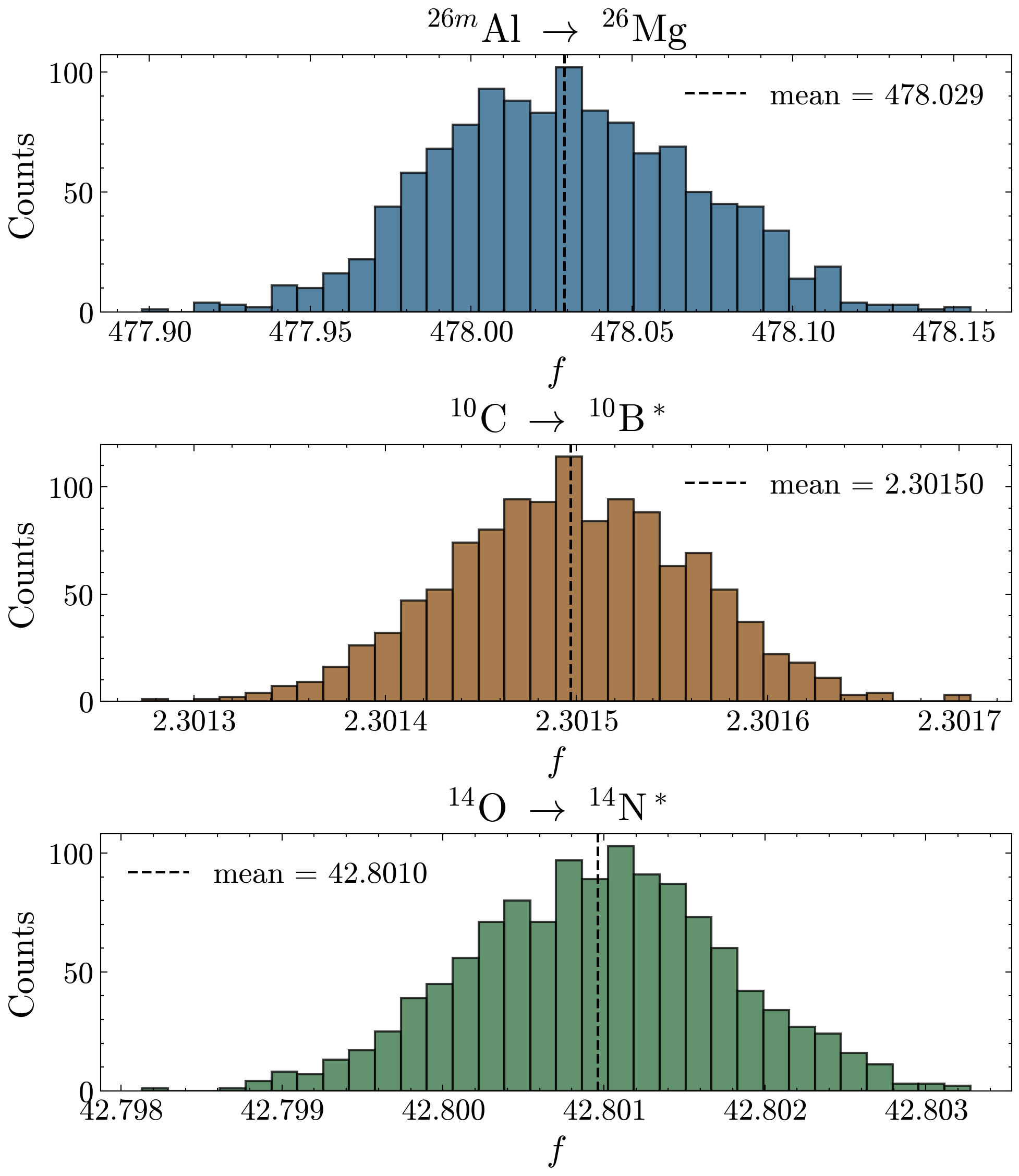}
	\caption{Histograms of sampled statistical rate functions $f$ for the three transitions, obtained by propagating paired parent/daughter charge-density ensembles through the Coulomb-corrected phase-space calculation. Dashed lines indicate the sample
  means.}
	\label{fig:fsamples}
\end{figure}

\textbf{Data on nuclear charge radii} -- Our strategy requires an experimental nuclear charge radius, typically available only for the stable $T_z=+1$ member of an isotriplet. However, the $T_z=-1$ charge radius can be inferred from the mirror-shift fit~\cite{Ohayon:2024dwt,Novario:2021low}, while the isospin formula yields the $T_z=0$ charge radius. We adopt the compilation in Table~7 of Ref.~\cite{Ohayon:2024dwt}; the quoted uncertainties are comparable to those of the measured radii.

This procedure also enables a semi-empirical test of isospin symmetry by comparing the predicted $T_z=0$ radius to available experimental data. Refs.~\cite{Ohayon:2024dwt, 2025-K} find overall consistency with isospin symmetry, albeit with sizable uncertainties. A notable exception is $A=26$, where the predicted ${}^{26\mathrm{m}}\text{Al}$ charge radius 
%of 3.088(3)~fm 
differs from the measured value %3.132(8)~fm 
by $5\sigma$. This points towards a unique opportunity to probe large isospin breaking effects, achievable through an experimental measurement of the ${}^{26}\text{Si}$ charge radius. In this work, we adopt the experimental value of the ${}^{26\mathrm{m}}\text{Al}$ charge radius.

\textbf{$f$-computation and uncertainty quantification} -- Using the procedure above, we compute the statistical rate function $f$ for the superallowed decays of ${}^{10}\text{C}$, ${}^{14}\text{O}$, and ${}^{26\mathrm{m}}\text{Al}$. For each transition, we construct the charge densities of the $T_z=0,+1$ nuclei and obtain $\rho_{\text{cw}}$ from isospin symmetry, then follow Ref.~\cite{Gorchtein:2025wli} to evaluate $f$.

We estimate uncertainties from charge-density inputs with two approaches. \textbf{Method~1:} We fix radii at their central values and vary the Hamiltonians and IMSRG truncations used to compute the $V$-factors; the spread in $f$ quantifies the \textit{ab initio} (i.e., $V$-factor) uncertainty. With $V$-factors fixed, we then vary the radii between the central and maximum values. This provides a straightforward assessment of the impact of each individual source of uncertainty. \textbf{Method~2:} We randomly sample a large ensemble of charge densities by varying both radii and $V$-factors within their uncertainties, compute $f$ for each sample, and take the standard deviation of the resulting distribution (Fig.~\ref{fig:fsamples}) as the combined uncertainty
from the radii and \textit{ab initio} calculations. For both methods, we include common uncertainties from $Q_{\text{EC}}$ and atomic screening~\footnote{The uncertainty of the atomic screening correction was taken to be 10\% of its actual size, following the treatment in Ref.~\cite{Seng:2023cgl}. This is an extremely conservative estimate and should by no means hinder future efforts to reduce uncertainties from other sources, such as improved measurements of $Q_\text{EC}$ values.}. The outcomes from the two methods are in good agreement, and we quote the final results based on the more sophisticated Method 2 as follows:
\begin{eqnarray}
{}^{10}\text{C}\rightarrow {}^{10}\text{B}^*&:& f=2.30150(71)_{Q_{\text{EC}}}(43)_{\text{scr}}(6)_{\text{den}}\nonumber\\
{}^{14}\text{O}\rightarrow {}^{14}\text{N}^*&:& f=42.8010(77)_{Q_{\text{EC}}}(63)_{\text{scr}}(8)_{\text{den}}\nonumber\\
{}^{26\mathrm{m}}\text{Al}\rightarrow {}^{26}\text{Mg}&:& f=478.029(101)_{Q_{\text{EC}}}(82)_{\text{scr}}(40)_{\text{den}}\,,\nonumber
\end{eqnarray}
where ``den'' is the charge density uncertainty, dominated by the charge radii. The $V$-factor contribution is $\mathcal{O}(0.001\%)$ or smaller, demonstrating that our procedure tightly constrains nuclear-size effects. Overall, we find a downward shift at the level of $\sim 0.01\%$ relative to the traditional evaluation~\cite{Hardy:2020qwl}, which did not incorporate the most precise charge density inputs or the correlation between $\rho_{\text{ch}}$ and $\rho_{\text{cw}}$.

\begin{figure}
	\centering
	\includegraphics[width=0.90\columnwidth]{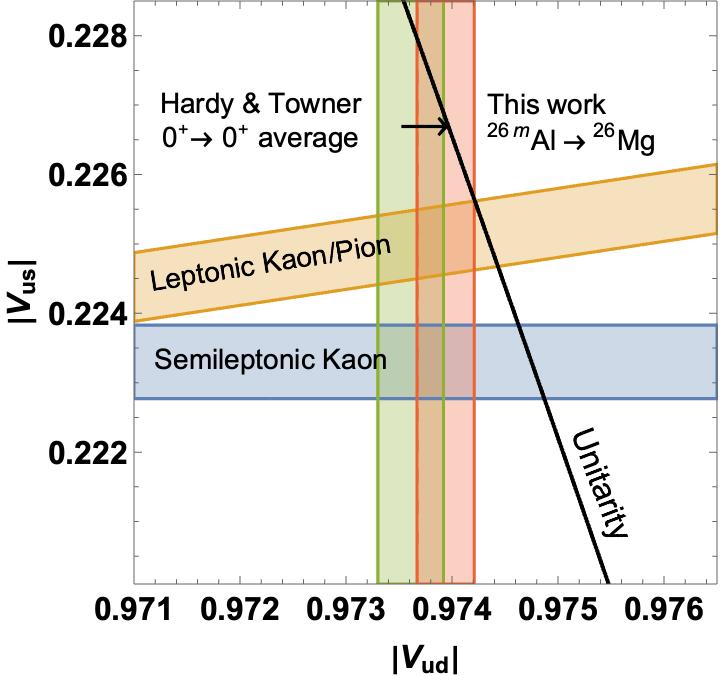}
	\caption{Updated plot of various $V_{ud}$ and $V_{us}$ determinations. The red vertical band represents our new result, $|V_{ud}|=0.97394(27)$, using %$\mathcal{F}t=3070.16(0.85)(0.20)(1.44)$ 
    $\mathcal{F}t=3070.16(1.68)$ 
    from $^{26\mathrm{m}}\text{Al}$ decay. The green vertical band, $|V_{ud}|=0.97373(31)$, is taken from Ref.~\cite{Hardy:2020qwl}. The black line represents the unitarity condition $\Delta_\text{CKM}=0$.}
	\label{fig:VudVus}
\end{figure}

\textbf{Implications for $V_{ud}$} -- We study the impact of our work on $V_{ud}$ computed via~\cite{Hardy:2020qwl}:
\begin{equation}\label{eq:Vud}
|V_{ud}|^2= \frac{2984.431(3)\,\text{s}}{\mathcal{F}t(1+\Delta_R^V)}~,
\end{equation}
using the transition ${}^{26\mathrm{m}}\text{Al}\rightarrow {}^{26}\text{Mg}$, which alone returns a $V_{ud}$ as precise as the global determination~\cite{Gorchtein:2025wli}. $\Delta_R^V$ is a nucleus-independent radiative correction that we take from Ref.~\cite{Gorchtein:2023srs}. We further account for recent studies of long-distance radiative corrections $\delta_R'$ at the $\mathcal{O}(Z\alpha^2)$ order in the EFT framework~\cite{Crosas:2025xyv,Cao:2025zxs}, which increases the $\mathcal{F}t$-value of this transition by 0.019\%.

Figure~\ref{fig:VudVus} compares the value of $V_{ud}$ obtained from $^{26\mathrm{m}}\text{Al}$ alone with that from the full set of superallowed decays, together with $V_{us}$ and $V_{us}/V_{ud}$ extracted from kaon and pion decays~\cite{Cirigliano:2022yyo}. Our updated value of $|V_{ud}|$ from $^{26\mathrm{m}}\text{Al}$ implies $\Delta_\text{CKM}=\{-1.58(58),-0.79(60)\}$ (in units of $10^{-3}$) when combined with $V_{us}$ from semileptonic and leptonic kaon decays, respectively~\cite{Cirigliano:2022yyo}. The corresponding unitarity tensions are 2.7 and 1.3$\sigma$. This should be compared with $\{-1.70(66),-0.93(69)\}$ from our previous work on $^{26\mathrm{m}}\text{Al}$ (corrected for the EFT shift) that did not include an \textit{ab initio} calculation~\cite{Gorchtein:2025wli}, for which the unitarity tensions were 2.6 and 1.3$\sigma$. We find that the central values of $\Delta_\text{CKM}$ decrease, while the substantially improved theoretical precision leaves the significance of the unitarity tension largely unchanged. This illustrates how a comprehensive treatment of nuclear-size effects can place the interpretation of new-physics searches in low-energy processes on a more solid footing.

\textbf{Discussions and outlook} -- For SM tests using superallowed $\beta$ decays it is crucial to treat nuclear finite size accurately. Here, we have presented a systematic approach that combines charge radii data, mirror-shift parameterization, isospin symmetry, and \textit{ab initio} charge form factors. We have shown that \textit{ab initio} calculations predict dimensionless moment ratios (the $V$-factors) with high precision compared with absolute moments, and that $f$ becomes essentially model independent once $\langle r^{2}\rangle$, $V_{24}$, and $V_{26}$ are fixed. For illustration, we have computed $f$ for ${}^{10}\text{C}\!\rightarrow {}^{10}\text{B}^*$, ${}^{14}\text{O}\!\rightarrow {}^{14}\text{N}^*$, and ${}^{26\mathrm{m}}\text{Al}\!\rightarrow {}^{26}\text{Mg}$ with a rigorous uncertainty budget that substantially strengthens the test for the first-row CKM unitarity. Our prescription applies to 21 of the 23 measured superallowed transitions; the exceptions are ${}^{66}\text{As}\!\rightarrow{}^{66}\text{Ge}$ and ${}^{70}\text{Br}\!\rightarrow{}^{70}\text{Se}$, where no isotriplet charge radii are available. A full re-evaluation of $f$ for the remaining transitions will be presented in follow-up work.

These results open several directions for future precision studies. First, the demonstrated stability of \textit{ab initio} $V$-factor predictions suggests that they could provide key theory input for extracting charge radii from muonic atom energy levels~\cite{Angeli:2026dlg}, further improving determinations of $f$ and fostering collaboration between nuclear \textit{ab initio} and atomic physics communities. Second, because isospin symmetry is central to our framework, it should be tested experimentally, particularly in the $A=26$ isotriplet that is pivotal for $V_{ud}$; semi-empirical analyses predict sizable isospin breaking~\cite{Ohayon:2024dwt}, which could be probed by measuring the $^{26}\text{Si}$ charge radius. Finally, our approach can complement the growing EFT program for electromagnetic effects in $\beta$ decays~\cite{Hill:2023acw,Hill:2023bfh,Borah:2024ghn,Crosas:2025xyv,Cao:2025zxs} by helping constrain unknown low-energy constants in higher-order operators.

\begin{acknowledgements}

\textbf{Acknowledgments} -- We thank Bob Wiringa, Garrett King and Saori Pastore for many useful discussions. BCH thanks Ragnar Stroberg for useful discussions and for facilitating access to computational resources at the Notre Dame Center for Research Computing.
BCH acknowledges support from the National Science Foundation (NSF) FRHTP program under award No.~PHY-2402275. The work of LP was supported by the National Science Foundation under Grant No.~PHY-2412612) and the US Department of Energy (Contract No.~DE-AC05-00OR22725). 
The work of CYS is supported in part by the U.S. Department of Energy Topical Collaboration ``Nuclear Theory for New Physics," award No.~DE-SC0023663, and by the University of Tennessee, Knoxville. 
The work of MH was supported 
by the Laboratory Directed Research and Development Program of Oak Ridge National Laboratory, managed by UT-Battelle, LLC, for the U.S.\ Department of Energy and by the U.S.\ Department of Energy, Office of Science, Office of Advanced Scientific Computing Research and Office of Nuclear Physics, Scientific Discovery through Advanced Computing (SciDAC) program (SciDAC-5 NUCLEI).
M.G.~acknowledges support  by the Deutsche Forschungsgemeinschaft (DFG) - GO 2604/3-2, Projektnummer 495329596, and by the Cluster of Excellence ``Precision Physics, Fundamental Interactions, and Structure of Matter" (PRISMA$^{++}$ EXC 2118/2) funded by the German Research Foundation (DFG) within the German Excellence Strategy (Project ID 390831469).
This research used resources of the Oak Ridge Leadership Computing Facility located at Oak Ridge National Laboratory, which is supported by the Office of Science of the Department of Energy under contract No.~DE-AC05-00OR22725.

\end{acknowledgements}

	\bibliography{ref.bib}

\clearpage

\setcounter{page}{1}
\renewcommand{\thepage}{Supplementary Information -- S\arabic{page}}
\setcounter{table}{0}
\renewcommand{\thetable}{S\,\Roman{table}}
\setcounter{equation}{0}
\renewcommand{\theequation}{S\,\arabic{equation}}
\setcounter{figure}{0}
\renewcommand{\thefigure}{S\,\arabic{figure}}

\section{Supplementary Information}

\subsection{Details of IMSRG calculations}

In this section, we provide the technical details of the IMSRG method, including its setup and the calculation of nuclear charge form factors. 

\subsubsection{Method}
We use VS-IMSRG~\cite{Hergert2016,Stroberg2019} to solve the many-body Schr\"odinger equation. This method solves for a unitary transformation of a given Hamiltonian $H$ to decouple subspaces of the full $A$-body Hilbert space.
Specifically, it decouples a core and valence space from the rest of the Hilbert space,
yielding effective interactions and operators that can be solved using shell model techniques.

The similarity renormalization group transforms the Hamiltonian like
\begin{equation}\label{eq:Hsdef}
    H(s) \equiv U(s) H U^{\dagger}(s),
\end{equation}
where $U(s)$ is the unitary transformation as a function of the flow parameter $s$, with $U(0)=1$.
The aim is to obtain a unitary transformation such that as $s$ increases the off-diagonal piece is suppressed, $H^{\rm od}(s) \to 0$, where ``off-diagonal'' refers to the part of the Hamiltonian to be decoupled.
To achieve this, we specify an anti-Hermitian generator $\eta$ that evolves the transformation in $s$,
\begin{equation}\label{eq:dUds}
    \frac{d}{ds}U(s) = \eta(s)U(s).
\end{equation}
There is some freedom in choosing the form of $\eta$, and several choices have been used in the literature. Here we use the arctangent variant of the White generator~\cite{Hergert2016, White2002}.
Applying \eqref{eq:dUds} to \eqref{eq:Hsdef} we obtain a flow equation for the Hamiltonian 
\begin{equation}\label{eq:dHds}
    \frac{d}{ds}H(s) = [\eta(s),H(s)].
\end{equation}
Integrating the differential equation~\eqref{eq:dHds} yields $H(s)$.

The IMSRG transformation induces many-body interactions and operators, which must be truncated due to their computational cost.
There are two approximations of the IMSRG equations used in this work, the VS-IMSRG(2) and the VS-IMSRG(3f$_2$)~\cite{PhysRevC.110.044317}.
In the IMSRG(2) approximation all operators are truncated at the normal-ordered two-body level.
This approximation may be systematically improved by going to the IMSRG(3) approximation, including normal-ordered three-body operators~\cite{Heinz2021}.
The IMSRG(3) is, however, too computationally expensive to be practical currently.
As a result several approximations of the IMSRG(3) have been developed~\cite{Heinz2021, PhysRevC.110.044316, PhysRevC.110.044317, PhysRevC.111.034311},
which seek to capture the leading effects of normal-ordered three-body operators in approximate ways.
The IMSRG(3f$_2$) does this using a modified ansatz and a factorized evaluation of the IMSRG transformation~\cite{PhysRevC.110.044317}, allowing it to capture leading IMSRG(3) effects at the same computational cost as the IMSRG(2).
%  Due to the computational cost, a direct application of IMSRG(3) is impractical. Moreover, not all topologies contributing to the full IMSRG(3) commutator are equally important. 
% A straightforward truncation scheme is therefore to retain only terms that scale as $N^7$ or less.
% This truncation scheme, which we denote IMSRG(3n7), was previously explored in Refs.~\cite{Heinz2021,PhysRevC.110.044316}.
% IMSRG(3f$_2$) is a further truncated version of IMSRG(3n7) that effectively captures the most important three-body correlations with lower computational effort.

Instead of directly integrating the flow equation~\eqref{eq:dHds}, we solve the IMSRG equations using the Magnus expansion~\cite{Morris2015PRC_Magnus}. The unitary transformation is expressed as the exponentiation of the anti-Hermitian Magnus operator $\Omega$
\begin{align} \label{eq:Omega}
U(s) = e^{\Omega(s)}\,.
\end{align} 
When the norm of the Magnus operator exceeds a given threshold, it is advantageous to split the transformation into two smaller unitary transformations. Consequently, Eq.~\eqref{eq:Omega} becomes
\begin{align} \label{eq:Multi-Omega}
U(s) = U(s - s_1) U(s_1)  = e^{\Omega(s - s_1)}  e^{\Omega(s_1)}\,.
\end{align}
Our IMSRG(2) calculations employ such a splitting approach, which can help reduce numerical errors~\cite{n348-lvll},
while our IMSRG(3f$_2$) do not~\cite{PhysRevC.110.044317}.
The robustness of our $V$-factor predictions is clearly not sensitive to this choice.

\subsubsection{Basic setup}

We expand our calculations in a spherical harmonic-oscillator (HO) single-particle basis with $\hbar\omega=16~\mathrm{MeV}$.
The basis is truncated according to $e=2n+\ell \leq e_{\mathrm{max}} = 12$, where $n$ and $\ell$ denote the radial and orbital angular-momentum quantum numbers.
%To compute the expectation value of the charge form factor operator in Eqs.\eqref{eq:Fch},\eqref{eq:Fchint}, we employ a single-particle basis of spherical harmonic oscillator eigenstates of frequency $\hbar\omega$ = 16 MeV, truncated according to $e=2n+\ell \leq e_{\text{max}}$, where $n$ and $\ell$ are the radial and angular momentum quantum numbers, respectively.
Three-particle states used in the matrix elements of three-body operators are further truncated by $e_1+e_2+e_3\leq E_{\text{3max}} = 18$~\cite{Miyagi2023EPJA_NuHamil}.
% Both IMSRG(2) and IMSRG(3f$_2$) are performed with $e_{\text{max}}=12$ and $E_{\text{3max}}=18$.
We employ an optimized natural orbital basis following the prescription of Ref.~\cite{PhysRevC.103.014321}.
% Unless otherwise indicated, we employ the natural orbital basis to speed up the convergence~\cite{PhysRevC.103.014321}.
We solve for the natural orbital basis in the full HO single-particle basis with $e_{\max}=12$,
then transform all operators to the new basis,
and then truncate to a smaller model space with $e_{\max}^\mathrm{NAT}=10$ before solving the IMSRG.
This reduces the computational cost of the many-body calculation while delivering results that are well converged with respect to the model-space truncation.
% To reduce the computational cost, we introduce an additional basis truncation. 
% After transforming the interactions and operators normal ordered with respect to the reference state, we truncate the model space to $e_{\max}=10$ before performing the IMSRG evolution.
% The charge form factor $F_{\mathrm{ch}}(q^2)$ is evaluated at momentum transfers $q$ up to 1~$\mathrm{GeV}$.
% for the IMSRG(2) calculations. For the IMSRG(3f$_2$) calculations, it is evaluated up to $q=190$~$\mathrm{MeV}$.
% We interpolate $F_{\mathrm{ch}}(q^2)$ using the Gaussian-process (GP) regression, following Ref.~\cite{MIYAGI2026140032}.

We use the $p$-shell valence space for our computations of the moments of ${}^{10}\mathrm{C}$, ${}^{10}\mathrm{B}^*$, ${}^{14}\mathrm{O}$, and ${}^{14}\mathrm{N}^*$, and we use an $sd$-shell valence space for our computations of the moments of ${}^{26\mathrm{m}}\mathrm{Al}$ and ${}^{26}\mathrm{Mg}$.
An estimate of the reference-state dependence is not included in this work. 
Throughout, we use the parent nucleus as the reference state.

\begin{table}[t]
\centering
\renewcommand{\arraystretch}{1.1}
\begin{ruledtabular}
\begin{tabular}{c c c c c c}
$e_{\max}^\mathrm{NAT}$ 
& $\langle r^{2}\rangle$ 
& $\langle r^{4}\rangle$ 
& $\langle r^{6}\rangle$ 
& $V_{24}$ 
& $V_{26}$ \\
\\[-2.5ex] \hline & \\[-2.5ex]
$8$  & $8.6767(1)$ & $111.41(3)$ & $1863(7)$ & $0.90666(6)$ & $0.8397(5)$ \\
$10$ & $8.7046(1)$ & $112.45(3)$ & $1900(7)$ & $0.90601(6)$ & $0.8384(5)$ \\
$12$ & $8.7393(1)$ & $113.86(3)$ & $1955(8)$ & $0.90499(6)$ & $0.8360(6)$ 
\end{tabular}
\end{ruledtabular}
\caption{
The model-space size $e_{\max}^\mathrm{NAT}$ dependence of the charge-radius moments and $V$-factors for $^{26\mathrm{m}}\mathrm{Al}$ using 1.8/2.0~(EM) with $E_{3\mathrm{max}}=18$ and $\hbar\omega=16~\mathrm{MeV}$.
The uncertainties of the moments are obtained from the GP, and the uncertainties of the $V$ factors are estimated by propagating the moment uncertainties assuming they are independent.
%The final calculations are performed at $e_{\max}=10$.
%, and the differences are reported with respect to the $e_{\max}=12$ results.
}
\label{tab:emax_convergence}
\end{table}

\subsubsection{Computation of $F_\mathrm{ch}(q^2)$}

We compute the translationally invariant charge form factor from the expectation value of the charge form-factor operator in Eq.~\eqref{eq:Fch} and by restoring translational invariance following Eq.~\eqref{eq:Fchint}.
The charge form factor $F_{\mathrm{ch}}(q^2)$ is evaluated at a set of momentum transfers $q$. For the IMSRG(2) calculations, we use values up to $q=1040~\mathrm{MeV}$, while for the IMSRG(3f$_2$) calculations we use values
up to $q=190~\mathrm{MeV}$. 
Following Ref.~\cite{MIYAGI2026140032}, we use Gaussian-process (GP) regression to obtain a smooth representation of $F_{\mathrm{ch}}(q^2)$ and evaluate the derivatives analytically. 
The lowest 25 momentum-transfer points, corresponding to approximately $q_{\max}=0.8~\mathrm{fm^{-1}}$, are used as the training data for the GP.

More specifically, for the IMSRG(2) calculations, the form factor is evaluated at $q = 0$, 1, 2, 3, 4, 5, 10, 15, 20, 25, 30, 35, 40, 45, 50,
60, 70, 80, 90, 100, 115, 130, 145, 160, 175, 190,
240, 290, 340, 390, 440, 490, 540, 590, 640, 690,
740, 790, 840, 890, 940, 990, 1040~$\mathrm{MeV}$.
For the IMSRG(3f$_2$) calculations, it is evaluated at
$q = 0$, 1, 2, 3, 4, 5, 10, 15, 20, 25, 30, 35, 40, 45, 50, 60, 70, 80, 90, 100, 115, 130, 145, 160, 175, 190~$\mathrm{MeV}$.
The interpolation uncertainty in the extracted $V$ factors as quantified by the GP is found to be negligible.
For example, for $^{26m}$Al with the 1.8/2.0~(EM) interaction and a model-space truncation of $e^{NAT}_{\max}=10$, we obtain
\begin{eqnarray}
V_{24}=0.9060142 \pm 6.165\times 10^{-5},\nonumber\\
V_{26}=0.8383602 \pm 5.377\times 10^{-4}.\nonumber
\end{eqnarray}
Therefore, in the final uncertainty estimation, we neglect the GP contribution to the uncertainty of the \textit{ab initio} $V$ factors.

To estimate the residual uncertainty associated with the basis truncation, we study the dependence on the final basis truncation $e_{\max}^\mathrm{NAT}$.
We perform the IMSRG calculations in truncated spaces with $e_{\max}^\mathrm{NAT}=8$, 10, and $12$, while keeping $E_{3\mathrm{max}}=18$ and $\hbar\omega=16~\mathrm{MeV}$ fixed.
The final results presented in this work are obtained with $e_{\max}^\mathrm{NAT}=10$. 
As shown in Table~\ref{tab:emax_convergence}, increasing the model space from $e_{\max}^\mathrm{NAT}=10$ to $e_{\max}^\mathrm{NAT}=12$ changes $\langle r^{2}\rangle$, $\langle r^{4}\rangle$, and $\langle r^{6}\rangle$ by approximately $0.40\%$, $1.26\%$, and $2.90\%$, respectively. 
The dimensionless ratios are more stable, and $V_{24}$ and $V_{26}$ change by only $0.11\%$ and $0.27\%$, respectively, from $e_{\max}^\mathrm{NAT}=10$ to $e_{\max}^\mathrm{NAT}=12$. 
These results support the use of $e_{\max}^\mathrm{NAT}=10$ for the calculations reported in this work.

\subsection{Details of uncertainty quantification}

As stated in the main text, we adopt two different methods to assess the uncertainties in $f$ from different sources. Details of each method are presented as follows.

\subsubsection{Method 1}

\begin{table*}
\renewcommand{\arraystretch}{1.3}
\begin{ruledtabular}
\begin{tabular}{l l r r r r r r r r}
Method & Hamiltonian & $V^{0}_{24}$ & $V^{0}_{26}$ & $V^{+}_{24}$ & $V^{+}_{26}$ & $r_{0}$ (fm) & $r_{+}$ (fm) & $Q_{\text{EC}}$ (MeV) & $f$ \\
\hline 
\hline 
\multirow{3}{*}{VS-IMSRG(2)} & \multirow{3}{*}{1.8/2.0~(EM)} & \multirow{3}{*}{0.877059} & \multirow{3}{*}{0.782291} & \multirow{3}{*}{0.890526} & \multirow{3}{*}{0.81216} & 2.531 & 2.361 & \textbf{1.908061} & 2.30231\tabularnewline
\cline{7-10}
 &  &  &  &  &  & \textbf{2.569} & 2.361 & 1.907994 & 2.30154\tabularnewline
\cline{7-10}
 &  &  &  &  &  & 2.531 & \textbf{2.397} & 1.907994 & 2.30165\tabularnewline
\hline
 \multirow{3}{*}{VS-IMSRG(2)}& 1.8/2.0~(EM) &  0.877059&  0.782291& 0.890526 & 0.81216 & \multirow{3}{*}{2.531} & \multirow{3}{*}{2.361} & \multirow{3}{*}{1.907994} & 2.30160\tabularnewline
\cline{2-6}\cline{10-10}
 & 1.8/2.0~(EM7.5) & 0.875084 & 0.780299 & 0.888204 & 0.809547 &  &  &  & 2.30160\tabularnewline
\cline{2-6}\cline{10-10}
 & $\Delta$NNLO$_\mathrm{GO}$ & 0.87618 & 0.780693 & 0.88906 & 0.807761 &  &  &  & 2.30160\tabularnewline
\hline
\multirow{3}{*}{VS-IMSRG(3f$_2$)} & 1.8/2.0~(EM) & 0.883212 & 0.795782 & 0.894602 & 0.820906 & \multirow{3}{*}{2.531} & \multirow{3}{*}{2.361} & \multirow{3}{*}{1.907994}  & 2.30159\tabularnewline
\cline{2-6}\cline{10-10}
 & 1.8/2.0~(EM7.5) & 0.876779 & 0.786487 & 0.887091 & 0.808138 &  &  &  & 2.30160\tabularnewline
\cline{2-6}\cline{10-10}
 & $\Delta$NNLO$_\mathrm{GO}$ & 0.88528 & 0.799844 & 0.894687 & 0.819654 &  &  &  & 2.30159
% Mean (standard dev.) ...
\end{tabular}
\end{ruledtabular}
\caption{\label{tab:10Cresult}Predicted statistical rate functions $f$ for ${}^{10}\text{C}\rightarrow {}^{10}\text{B}^*$. In the top three rows, we keep the \textit{ab initio} input fixed and vary the charge radius of the $T_z=0$ nucleus $r_0$, the charge radius of the $T_z=+1$ nucleus $r_+$, and the electron capture $Q$-value $Q_\mathrm{EC}$ between their central values and their maximum values (indicated in bold).
In the middle three rows, we consider VS-IMSRG(2) predictions for the $V$-factors in the $T_z=0$ nucleus, $V_{24}^0$ and $V_{26}^0$, and the $T_z=+1$ nucleus, $V_{24}^+$ and $V_{26}^+$, for three Hamiltonians from chiral EFT.
In the bottom three rows, we consider instead VS-IMSRG(3f$_2$) predictions for the $V$-factors.
\\Data input: $r_\text{ch}({}^{10}\text{B}^*)=2.531(38)\:\text{fm}$, $r_\text{ch}({}^{10}\text{Be})=2.361(36)\:\text{fm}$, $Q_\text{EC}=1.907994(67)\:\text{MeV}$. \\Average result (only individual uncertainties larger than $5\times 10^{-3}$\% are displayed):  $f=2.3016(7)_{Q_\text{EC}}(4)_\text{scr}$.
}
\par\end{table*}

\begin{table*}
\renewcommand{\arraystretch}{1.3}
\begin{ruledtabular}
\begin{tabular}{l l r r r r r r r r}
Method & Hamiltonian & $V^{0}_{24}$ & $V^{0}_{26}$ & $V^{+}_{24}$ & $V^{+}_{26}$ & $r_{0}$ (fm) & $r_{+}$ (fm) & $Q_{\text{EC}}$ (MeV) & $f$ \\
\hline 
\hline 
\multirow{3}{*}{VS-IMSRG(2)} & \multirow{3}{*}{1.8/2.0~(EM)} & \multirow{3}{*}{0.891785} & \multirow{3}{*}{0.811609} & \multirow{3}{*}{0.901454} & \multirow{3}{*}{0.833964} & 2.623 & 2.508 & \textbf{2.831619} & 42.8076\tabularnewline
\cline{7-10}\cline{9-10}
 &  &  &  &  &  & \textbf{2.632} & 2.508 & 2.831543 & 42.7992\tabularnewline
\cline{7-10}\cline{10-10}
 &  &  &  &  &  & 2.623 & \textbf{2.517} &  2.831543 & 42.8004\tabularnewline
\hline
 \multirow{3}{*}{VS-IMSRG(2)} & 1.8/2.0~(EM) & 0.891785 & 0.811609 & 0.901454 & 0.833964 &  \multirow{3}{*}{2.623} & \multirow{3}{*}{2.508} & \multirow{3}{*}{2.831543 } & 42.7999\tabularnewline
\cline{2-6}\cline{10-10}
 & 1.8/2.0~(EM7.5) & 0.89479 & 0.817011 & 0.90523 & 0.839298 &  &  &  & 42.7999\tabularnewline
\cline{2-6}\cline{10-10}
 & $\Delta$NNLO$_\mathrm{GO}$ & 0.89178 & 0.810852 & 0.902742 & 0.835454 &  &  &  & 42.8000\tabularnewline
\hline
\multirow{3}{*}{VS-IMSRG(3f$_2$)} & 1.8/2.0~(EM) & 0.892212 & 0.812461 & 0.901403 & 0.833825 & \multirow{3}{*}{2.623} & \multirow{3}{*}{2.508} & \multirow{3}{*}{2.831543 } & 42.7998\tabularnewline
\cline{2-6}\cline{10-10}
 & 1.8/2.0~(EM7.5) & 0.894959 & 0.817085 & 0.907859 & 0.844733 &  &  &  & 42.8002\tabularnewline
\cline{2-6}\cline{10-10}
 & $\Delta$NNLO$_\mathrm{GO}$ & 0.892575 & 0.813121 & 0.904046 & 0.837868 &  &  &  & 42.8001
\end{tabular}
\end{ruledtabular}
\caption{\label{tab:14Oresult}Same as Table~\ref{tab:10Cresult} but for  ${}^{14}\text{O}\rightarrow {}^{14}\text{N}^*$. \\
Data input: $r_\text{ch}({}^{14}\text{N}^*)=2.623(9)\:\text{fm}$, $r_\text{ch}({}^{14}\text{C})=2.508(9)\:\text{fm}$, $Q_\text{EC}=2.831543(76)\:\text{MeV}$. \\
Average result: $f=42.800(8)_{Q_\text{EC}}(6)_\text{scr}$.
}
\par\end{table*}

\begin{table*}
\renewcommand{\arraystretch}{1.3}
\begin{ruledtabular}
\begin{tabular}{l l r r r r r r r r}
Method & Hamiltonian & $V^{0}_{24}$ & $V^{0}_{26}$ & $V^{+}_{24}$ & $V^{+}_{26}$ & $r_{0}$ (fm) & $r_{+}$ (fm) & $Q_{\text{EC}}$ (MeV) & $f$ \\
\hline 
\hline 
\multirow{3}{*}{VS-IMSRG(2)} & \multirow{3}{*}{1.8/2.0~(EM)} & \multirow{3}{*}{0.90602} & \multirow{3}{*}{0.838353} & \multirow{3}{*}{0.907052} & \multirow{3}{*}{0.839918} & 3.132 & 3.030 & \textbf{4.23287}  & 478.128\tabularnewline
\cline{7-10}\cline{9-10}
 &  &  &  &  &  & \textbf{3.140} & 3.030 & 4.23272 & 478.002\tabularnewline
\cline{7-10}\cline{10-10}
 &  &  &  &  &  & 3.030 & \textbf{3.033} & 4.23272 & 478.034\tabularnewline
\cline{7-10}\cline{10-10}
\hline
 \multirow{3}{*}{VS-IMSRG(2)}& 1.8/2.0~(EM) & 0.90602 & 0.838353 & 0.907052 & 0.839918 &\multirow{3}{*}{3.132}  & \multirow{3}{*}{3.030} & \multirow{3}{*}{4.23272 } & 478.027\tabularnewline
\cline{2-6}\cline{10-10}
 & 1.8/2.0~(EM7.5) & 0.910733 & 0.846652 & 0.910906 & 0.846836 &  &  &  & 478.016\tabularnewline
\cline{2-6}\cline{10-10}
 & $\Delta$NNLO$_\mathrm{GO}$ & 0.908996 & 0.8426 & 0.909454 & 0.843408 &  &  &  & 478.019\tabularnewline
\hline
\multirow{3}{*}{VS-IMSRG(3f$_2$)} & 1.8/2.0~(EM) & 0.905932 & 0.839231 & 0.906857 & 0.840003 &  \multirow{3}{*}{3.132}  & \multirow{3}{*}{3.030} & \multirow{3}{*}{4.23272 } & 478.029\tabularnewline
\cline{2-6}\cline{10-10}
 & 1.8/2.0~(EM7.5) & 0.912871 & 0.851619 & 0.913008 & 0.851465 &  &  &  & 478.015\tabularnewline
\cline{2-6}\cline{10-10}
 & $\Delta$NNLO$_\mathrm{GO}$ & 0.909909 & 0.845447 & 0.910135 & 0.845083 &  &  &  & 478.019
\end{tabular}
\end{ruledtabular}
\caption{\label{tab:26mAlresult}
Same as Table~\ref{tab:10Cresult} but for ${}^{26\mathrm{m}}\text{Al}\rightarrow {}^{26}\text{Mg}$. \\ 
Data input: $r_\text{ch}({}^{26\mathrm{m}}\text{Al})=3.132(8)\:\text{fm}$, $r_\text{ch}({}^{26}\text{Mg})=3.030(3)\:\text{fm}$, $Q_\text{EC}=4.23272(15)\:\text{MeV}$. \\ Average result: $f=478.02(10)_{Q_\text{EC}}(8)_\text{scr}(3)_{r_\text{Al}}$.
}
\par\end{table*}

%\begin{table*}
%\begin{tabular}{|c|c|c|c|c|c|c|}
%\hline 
%\noalign{\vskip 0.5mm}
%\multirow{2}{*}{Interaction} & \multicolumn{2}{c|}{$f(\ ^{10}\text{C}\rightarrow\ ^{10}\text{B}^*)$} & \multicolumn{2}{c|}{$f(\ ^{14}\text{O}\rightarrow\ ^{14}\text{N}^*)$} & \multicolumn{2}{c|}{$f(\ ^{26\mathrm{m}}\text{Al}\rightarrow\ ^{26}\text{Mg})$}\tabularnewline
%\cline{2-7}
% & IMSRG(2) & IMSRG(2)+data & IMSRG(2) & IMSRG(2)+data & IMSRG(2) & IMSRG(2)+data\tabularnewline
%\hline 
%\hline 
%1.8/2.0~(EM) & 2.30167 & 2.30160 & 42.8012 & 42.7999 & 478.276 & 478.027\tabularnewline
%\hline 
%1.8/2.0~(EM7.5) & 2.30162 & 2.30160 & 42.7993 & 42.7999 & 478.210 & 478.016\tabularnewline
%\hline 
%$\Delta$NNLO$_\mathrm{GO}$ & 2.30167 & 2.30160 & 42.7992 & 42.8000 & 478.261 & 478.019\tabularnewline
%\hline 
% Spread %& 0.00x & 0.00...
%\end{tabular}
%\caption{Comparison of interaction-dependence of $f$ between direct \textit{ab initio} evaluation (i.e. using directly the \textit{ab initio} charge form factors) and \textit{ab initio}+data (i.e. combining experimental $\langle r^2\rangle$ and \textit{ab initio} $V$-factors). }
%\par\end{table*}

In the first method, we parameterize all charge densities using the 3pF model:
\begin{equation}
        \rho_\text{ch}(r)=\frac{\rho_0(1+wr^2/c^2)}{1+\exp\{(r^2-c^2)/a^2\}}~,
\end{equation}
with the parameters $a$, $c$, and $w$ fixed to reproduce the given values of $\langle r^2\rangle$, $V_{24}$ and $V_{26}$. As we demonstrated in the main text, different choices of model only affect the central value at the 0.001\% level, which is below our precision goal.

To assess the uncertainty due to the \textit{ab initio} calculation, we fix $\langle r^2\rangle$, $Q_\text{EC}$ and compute $f$ using the $V$-factors resulting from different IMSRG truncations and different Hamiltonians. Meanwhile, to assess the uncertainty due to charge radii and $Q_\text{EC}$, we compute $f$ by varying these parameters between their central value and maximum value, one at a time, under the same \textit{ab initio} setting [VS-IMSRG(2) with the 1.8/2.0~(EM) interaction]. The results are summarized in Table~\ref{tab:10Cresult}, \ref{tab:14Oresult}, \ref{tab:26mAlresult}, from which one clearly observes that the variations of the outcome due to different \textit{ab initio} settings are on the order of 0.001\% or smaller. This is a strong demonstration of the robustness of our proposed prescription in pinning down finite size effects in $f$. 

\subsubsection{Method 2}

\begin{figure}[h]
	\centering
	\includegraphics[width=0.9\columnwidth]
    {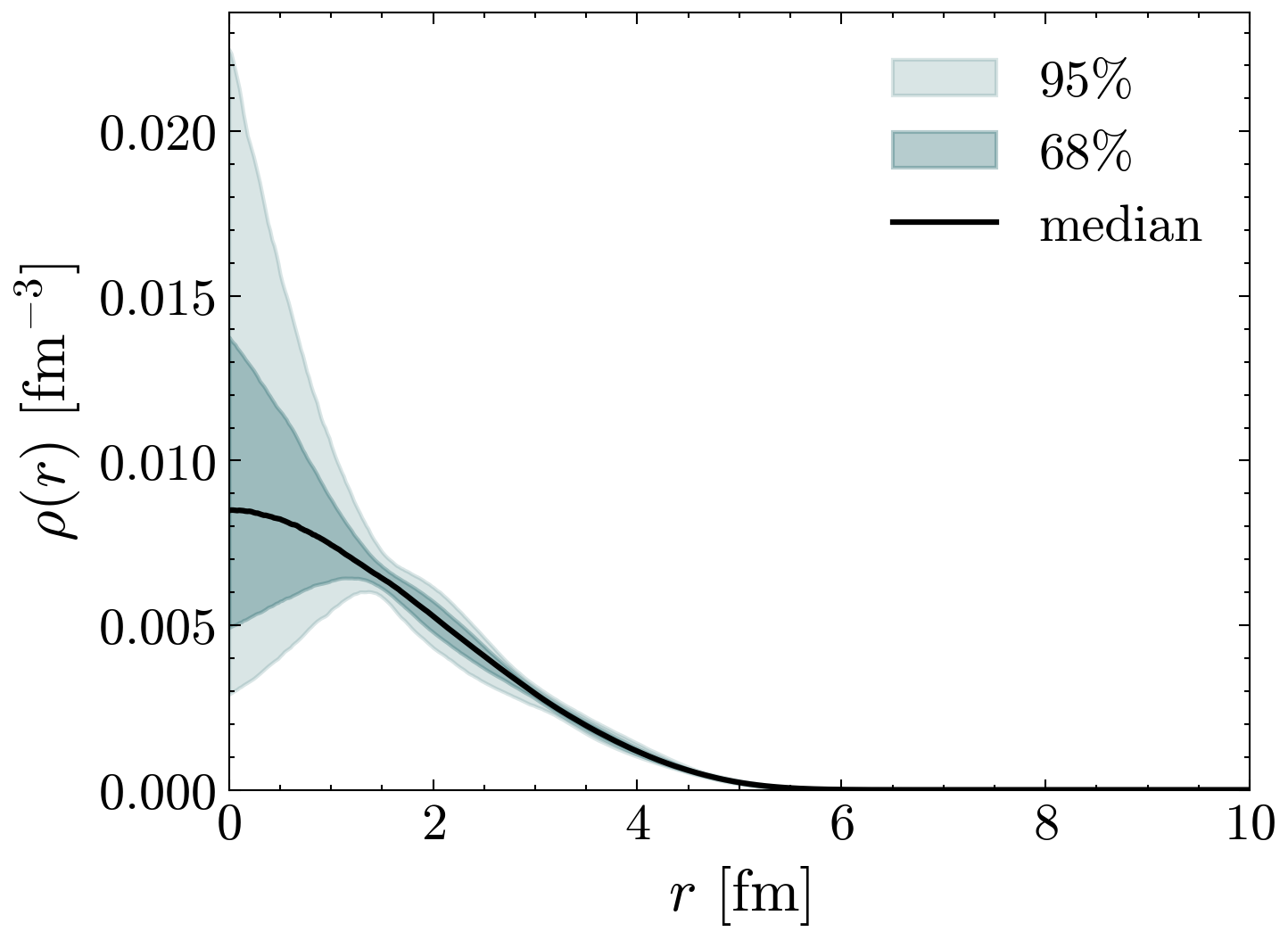}
	\caption{The combined plot of all ${}^{26\mathrm{m}}\text{Al}$ charge densities generated in Method 2.}
	\label{fig:al26density}
\end{figure}

\begin{table}[t]
\begin{ruledtabular}
  \begin{tabular}{lccc}
  Transition & Mean $f$ & Std.~dev. & 95\% interval \\
  \midrule
  $^{26\mathrm{m}}$Al $\rightarrow$ $^{26}$Mg & 478.029 & 0.040 & [477.954, 478.107] \\
  $^{10}$C $\rightarrow$ $^{10}$B$^*$ & 2.30150 & 0.00006 & [2.30137, 2.30162] \\
  $^{14}$O $\rightarrow$ $^{14}$N$^*$ & 42.8010 & 0.0008 & [42.7994, 42.8025] 
  \end{tabular}  
\end{ruledtabular}
\centering
  \caption{\label{tab:method2result}Summary of sampled statistical rate functions. Intervals are empirical central probability intervals over the density ensemble.}
\end{table}

In the second method, we start with the IMSRG form factor results for the different interactions and obtain the $V$-factors using a Gaussian process. Then, we create a family of functions by exponentiating a set of radial Gaussians
\begin{align}
    g(r) = \sum_a c_a \exp(-(r-R_a)^2/s_a^2)\,,
\end{align}
whose coefficients $c_a$ are drawn randomly from a normal distribution. The widths are constant $s_a = 1$~fm and the centers $R_a$ are evenly spaced between the origin and 12~fm.
We exponentiate these function to obtain positive definite base densities
\begin{align}
    \rho_{\rm base}(r) = N_{\rm base} \exp(g(r)) R(r)\,,
\end{align}
where $N_{\rm base}$ is a normalization factor ensuring that $\rho_{\rm base}$ is normalized to one, and $R(r)$ is a regulator function that smoothly cuts off the density at a large distance.

We then use an exponential factor with Lagrange multipliers $\lambda_2, \lambda_4$ and $\lambda_6$ to create from this family of base densities $\rho_{\rm base}$ new densities that return the correct charge radius and moment ratios. Specifically,
\begin{align}
    \rho_\lambda(r) =N_\lambda \rho_{\rm base} \exp(\lambda_2 r^2 + \lambda_4 r^4 + \lambda_6)\,,
\end{align}
where $N_\lambda$ is again a normalization factor. The $\lambda$s are obtained through a nonlinear least squares fit. 

Figure~\ref{fig:al26density} provides an illustration of the family of charge densities generated with the method above. We compute values of $f$ by sampling the densities that were generated this way. This leads to the histogram in Fig.~\ref{fig:fsamples}. The final numerical result is given in Table~\ref{tab:method2result}, where the standard deviation represents collectively the uncertainty from the charge radii, $V$-factors and the density models. We find excellent agreement between the two methods, in both the central values and uncertainties.

\end{document}